\documentclass[12pt,reqno]{article}
\usepackage{amsfonts}
\usepackage{amsmath}
\usepackage{amsbsy}

\usepackage{amssymb,latexsym}
\numberwithin{equation}{section}

\begin{document}
 \allowdisplaybreaks[1]
\title{Dualisation of the D=9 Matter Coupled Supergravity}
\author{Nejat T. Y$\i$lmaz\\
Department of Mathematics
and Computer Science,\\
\c{C}ankaya University,\\
\"{O}\u{g}retmenler Cad. No:14,\quad  06530,\\
 Balgat, Ankara, Turkey.\\
          \texttt{ntyilmaz@cankaya.edu.tr}}
\maketitle
\begin{abstract}We perform the bosonic dualisation of the matter
coupled $\mathcal{N}=1$, $D=9$ supergravity. We derive the Lie
superalgebra which parameterizes the coset map whose Cartan form
realizes the second-order bosonic field equations. Following the
non-linear coset construction we present the first-order
formulation of the bosonic field equations as a twisted
self-duality condition.

\end{abstract}

\section{Introduction}
The dualisation of the bosonic sectors of the IIB supergravity
\cite{2B1,2B2,2B3} and the maximal supergravities which are
obtained from the $D=11$ supergravity \cite{d=11} by torodial
compactifications is performed in \cite{julia2}. The bosonic
dualisation which is another manifestation of the Lagrange
multiplier method \cite{pope} has enabled us to construct the
bosonic sectors of the above mentioned supergravity theories as
non-linear realisations. This can be considered as a
generalization of the non-linear structure of the scalar cosets of
these supergravity theories to include the bosonic fields as well.
The method of \cite{julia2} has also served to obtain the
first-order formulation of the bosonic sectors in which the
first-order field equations arise from a twisted self-duality
condition which the dualized Cartan form obeys. The scalar coset
manifolds of the supergravities studied in \cite{julia2} have
split real form global symmetry groups. The dualisation of a
generic scalar coset which has a split real form global symmetry
group is performed in \cite{nej1}. In \cite{nej2} this formulation
is generalized to the non-split scalar cosets \cite{ker1,ker2}.
The dualisation and the first-order formulation of the general
non-split scalar coset which is coupled to generic $m$-form matter
fields is performed in \cite{nej3}. In the light of \cite{nej2}
and \cite{nej3} the dualisation method of \cite{julia2} has been
extended to cover the class of supergravities which possess scalar
cosets based on non-split global symmetry groups and which contain
matter couplings. The dualisation and the non-linear realisation
of the $D=8$ and the $D=7$ matter coupled supergravities are
studied in \cite{nej4} and \cite{nej5} respectively.

In this work we present the non-linear realisation of the
$\mathcal{N}=1$, $D=9$ supergravity which is coupled to $N$ vector
multiplets \cite{d=9}. In section two we will introduce the matter
coupled $D=9$ supergravity and we will discuss the scalar coset
manifold. We will give the construction of the scalar Lagrangian
as a symmetric space sigma model
\cite{julia2,pope,nej1,nej2,ker1,ker2,julia1}. The second-order
bosonic field equations will also be derived in section two. In
section three we will derive the Lie superalgebra which
parameterizes the doubled coset element that generates the
dualized Cartan form. The construction will be based on the
requirement that the dualized Cartan form should yield the
second-order bosonic field equations obtained in section two via
the Cartan-Maurer equation. Finally we will explicitly calculate
the dualized Cartan form and obtain the first-order field
equations by using the twisted self-duality condition which the
dualized Cartan form satisfies \cite{julia2}.

\section{The $D=9$ Matter Coupled Supergravity}
The $\mathcal{N}=1$, $D=9$ supergravity is coupled to $N$ vector
multiplets in \cite{d=9}. The $\mathcal{N}=1$, $D=9$ supergravity
multiplet can be given as
\begin{equation}\label{21}
(e_{\mu}^{r},\psi_{\mu},\chi ,B_{\mu\nu},A_{\mu}, \sigma),
\end{equation}
where $e_{\mu}^{r}$ is the veilbein, $B_{\mu\nu}$ is a two-form
field, $A_{\mu}$ is a one-form field, $\sigma$ is the dilaton
whereas $\psi_{\mu}$ and $\chi$ are the spinor fields. The matter
coupled $D=9$ supergravity can be obtained by coupling \eqref{21}
an arbitrary number $N$ of $\mathcal{N}=1$ vector multiplets
\begin{equation}\label{22}
(\lambda ,A_{\mu},\varphi),
\end{equation}
where $\lambda$ is a spinor field, $A_{\mu}$ is a one-form and
$\varphi$ is the scalar. Thus the total field content of the $D=9$
matter coupled supergravity becomes
\begin{equation}\label{23}
(e_{\mu}^{r},\psi_{\mu},\chi ,B_{\mu\nu},A_{\mu}^{I},
\sigma,\lambda^{\alpha} ,\varphi^{\alpha}),
\end{equation}
where $\alpha=1,...,N$ and $I=1,...,N+1$. Our dualisation analysis
in the next section will be applied to the bosonic sector of this
theory therefore we are not interested in the nature of the spinor
fields.

The scalars $\varphi^{\alpha}$ for $\alpha=1,...,N$ parameterize
the scalar coset manifold $SO(N,1)/SO(N)$ where $SO(N,1)$ is in
general a non-compact real form of a semi-simple Lie group and
$SO(N)$ is its maximal compact subgroup. For this reason
$SO(N,1)/SO(N)$ is a Riemannian globally symmetric space for all
the $SO(N,1)$-invariant Riemannian structures on $SO(N,1)/SO(N)$
\cite{hel}. Therefore the scalar sector which consists of the
vector multiplet scalars of the $D=9$ matter coupled supergravity
can be formulated as a general symmetric space sigma model
\cite{nej2}. To construct the symmetric space sigma model
Lagrangian we will use the solvable Lie algebra parametrization
\cite{fre} for the parametrization of the scalar coset manifold
$SO(N,1)/SO(N)$. The solvable Lie algebra parametrization is a
consequence of the Iwasawa decomposition \cite{hel}
\begin{subequations}\label{24}
\begin{align}
so(N,1)&=\mathbf{k}_{0}\oplus \mathbf{s}_{0}\notag\\
\notag\\
&=\mathbf{k}_{0}\oplus \mathbf{h_{k}}\oplus
\mathbf{n_{k}},\tag{\ref{24}}
\end{align}
\end{subequations}
where $\mathbf{k}_{0}$ is the Lie algebra of $SO(N)$ and
$\mathbf{s}_{0}$ is the solvable Lie subalgebra of $so(N,1)$.
$\mathbf{h_{k}}$ is a subalgebra of the Cartan subalgebra
$\mathbf{h}_{0}$ of $so(N,1)$ which generates the maximal R-split
torus in $SO(N,1)$ \cite{nej2,ker2,hel}. The nilpotent Lie
subalgebra $\mathbf{n_{k}}$ of $so(N,1)$ is generated by a subset
$\{E_{m}\}$ of the positive root generators of $so(N,1)$ where
$m\in\Delta_{nc}^{+}$. The roots in $\Delta_{nc}^{+}$ are the
non-compact roots with respect to the Cartan involution $\theta$
induced by the Cartan decomposition \cite{nej2,nej3,hel}
\begin{equation}\label{25}
so(N,1)=\mathbf{k}_{0}\oplus\mathbf{u}_{0},
\end{equation}
where $\mathbf{u}_{0}$ is a vector subspace of $so(N,1)$. By using
the scalar fields $\varphi^{\alpha}$ of the coupling vector
multiplet and the generators of the solvable Lie algebra
$\mathbf{s}_{0}$ we can parameterize the representatives of the
scalar coset manifold $SO(N,1)/SO(N)$ as \cite{hel}
\begin{equation}\label{26}
L=\mathbf{g}_{H}\mathbf{g}_{N}=e^{\frac{1}{2}\phi
^{i}H_{i}}e^{\chi ^{m}E_{m}},
\end{equation}
where $\{H_{i}\}$ for $i=1,...,$ dim$\mathbf{h}_{k}$ are the
generators of $\mathbf{h}_{k}$ and $\{E_{m}\}$ for
$m\in\Delta_{nc}^{+}$ are the positive root generators which
generate $\mathbf{n_{k}}$. The scalars $\{\phi^{i}\}$ for
$i=1,...,$ dim$\mathbf{h}_{k}$ are called the dilatons and
$\{\chi^{m}\}$ for $m\in\Delta_{nc}^{+}$ are called the axions.
The coset representatives satisfy the defining equations of
$SO(N,1)$
\begin{equation}\label{27}
L^{T}\eta L=\eta\quad\quad,\quad\quad L^{-1}=\eta L^{T}\eta,
\end{equation}
where  $\eta=$diag$(-,+,+,+,..,+)$. Likewise in \cite{nej4} and
\cite{nej5} we will assume an $(N+1)$-dimensional representation
for the algebra $so(N,1)$ which generates symmetric matrix
representatives for the coset elements of \eqref{26}. Therefore
$L^{T}=L$ but as we have done in \cite{nej4} and \cite{nej5} we
will keep on using $L^{T}$ in our formulation for the sake of
accuracy. By using the definitions \eqref{26} and \eqref{27}, also
by using the fact that we have chosen the $(N+1)$-dimensional
representation for the algebra $so(N,1)$ such that $L^{T}=L$ we
have the following relations for the coset representatives under
the representation chosen
\begin{subequations}\label{28}
\begin{gather}
\partial_{i}L\equiv\frac{\partial L}{\partial\phi^{i}}=\frac{1}{2}H_{i}L\quad\quad,\quad\quad
\partial_{i}L^{T}\equiv\frac{\partial L^{T}}{\partial\phi^{i}}=\frac{1}{2}L^{T}H_{i}^{T},\notag\\
\notag\\
\partial_{i}L^{-1}\equiv\frac{\partial L^{-1}}{\partial\phi^{i}}=-\frac{1}{2}L^{-1}H_{i},\notag\\
\notag\\
(H_{i}L)^{T}=H_{i}L\quad\quad,\quad\quad H_{i}^{T}\eta=-\eta
H_{i}.\tag{\ref{28}}
\end{gather}
\end{subequations}
Now we will introduce the internal metric $\mathcal{M}$ as
\begin{equation}\label{29}
\mathcal{M}=L^{T}L.
\end{equation}
The scalar Lagrangian which governs the $N$ scalar fields of the
vector multiplet can be constructed as
\begin{equation}\label{30}
 \mathcal{L}_{scalar}=\frac{1}{16}tr(\ast d\mathcal{M}^{-1}\wedge
 d\mathcal{M}),
\end{equation}
which is the Lagrangian for the symmetric space sigma model
\cite{julia2,nej2,ker2,nej3}. Apart from gravity the bosonic
Lagrangian of the $D=9$ matter coupled supergravity can now be
given as \cite{d=9}
\begin{equation}\label{210}
\begin{aligned}
\mathcal{L}&=\frac{7}{4} \ast d\sigma\wedge
d\sigma+\frac{1}{2}e^{-4\sigma}\ast G
\wedge G\\
\\
&\quad +\frac{1}{16}tr( \ast d\mathcal{M}^{-1}\wedge
d\mathcal{M})-\frac{1}{2}e^{-2\sigma} F\wedge\mathcal{M} \ast F,
\end{aligned}
\end{equation}
where the coupling between the field strengths $F^{I}=dA^{I}$ for
$I=1,...,(N+1)$ and the scalars which parameterize the coset
$SO(N,1)/SO(N)$ can be explicitly written as
\begin{equation}\label{211}
-\frac{1}{2}e^{-2\sigma} F\wedge\mathcal{M}\ast
F=-\frac{1}{2}e^{-2\sigma}\mathcal{M}_{IJ} F^{I}\wedge \ast F^{J}.
\end{equation}
The Chern-Simons three-form $G$ is defined as \cite{d=9}
\begin{equation}\label{212}
 G=dB+\eta_{IJ}A^{I}\wedge F^{J}.
\end{equation}
We should remark one point here that the construction of
\cite{d=9} is built on a symmetric invariant second-rank tensor
$L_{IJ}$ of the group $SO(N,1)$ which is specified by the choice
of the $(N+1)$-dimensional representation for the coset elements
$L$ of $SO(N,1)/SO(N)$. In our formulation we have used the
solvable Lie algebra parametrization \eqref{26} whose images are
in the matrix group $SO(N,1)$ as a result of the exponential map
and the proposed fundamental representation of $so(N,1)$. Thus in
our assumed representation our coset representatives are elements
of the global symmetry group $SO(N,1)$. Therefore due to the
definition of $L_{IJ}$ in \cite{d=9} $L_{IJ}=\eta_{IJ}$ for our
general coset representation. This is in conformity with the
construction in \cite{nej4} and \cite{nej5}.

By varying the Lagrangian \eqref{210} with respect to the fields
$\sigma,B$ and $\{A^{I}\}$ we find the corresponding field
equations as
\begin{subequations}\label{213}
\begin{gather}
\frac{7}{2}d(\ast d\sigma)=-2e^{-4\sigma} \ast G\wedge
G+e^{-2\sigma}\mathcal{M}_{IJ} \ast F^{I}\wedge
F^{J},\notag\\
\notag\\
d(e^{-4\sigma}\ast G)=0,\notag\\
\notag\\
d(e^{-2\sigma}\mathcal{M}_{IJ}\ast
F^{J})=2e^{-4\sigma}\eta_{IJ}F^{J}\wedge\ast G.\tag{\ref{213}}
\end{gather}
\end{subequations}
In \cite{nej2} the Cartan-Maurer form
\begin{equation}\label{214}
\mathcal{G}_{0}=dLL^{-1},
\end{equation}
for a generic scalar coset is already calculated as
\begin{equation}\label{215}
\begin{aligned}
\mathcal{G}_{0}&=\frac{1}{2}d\phi ^{i}H_{i}+e^{%
\frac{1}{2}\alpha _{i}\phi ^{i}}U^{\alpha }E_{\alpha }\\
\\
&=\frac{1}{2}d\phi ^{i}H_{i}+\overset{\rightharpoonup }{
\mathbf{E}^{\prime }}\:\mathbf{\Omega }\:\overset{\rightharpoonup
}{d\chi }.
\end{aligned}
\end{equation}
Here we have introduced the column vector
\begin{equation}\label{216}
U^{\alpha}=\mathbf{\Omega}^{\alpha}_{\beta}d\chi^{\beta},
\end{equation}
and the row vector
\begin{equation}\label{217}
(\overset{\rightharpoonup }{
\mathbf{E}^{\prime }})_{\alpha}=e^{%
\frac{1}{2}\alpha _{i}\phi ^{i}}E_{\alpha},
\end{equation}
where $\mathbf{\Omega}$ is an
dim$\mathbf{n_{k}}\times$dim$\mathbf{n_{k}}$ matrix
\begin{equation}\label{217}
\begin{aligned}
 \mathbf{\Omega}&=\sum\limits_{m=0}^{\infty }\dfrac{\omega
^{m}}{(m+1)!}\\
\\
&=(e^{\omega}-I)\,\omega^{-1}.
\end{aligned}
\end{equation}
The matrix $\omega$ is
$\omega_{\beta}^{\gamma}=\chi^{\alpha}K_{\alpha\beta}^{\gamma}$.
The notation we use for the commutation relations of the solvable
Lie algebra $\mathbf{s}_{0}$ of $so(N,1)$ is as follows
\begin{equation}\label{218}
[H_{i},E_{\alpha}]=\alpha_{i}E_{\alpha}\quad ,\quad
[E_{\alpha},E_{\beta}]=K_{\alpha\beta}^{\gamma}E_{\gamma}.
\end{equation}
Since we have
\begin{equation}\label{219}
[E_ {\alpha},E_{\beta}]=N_{\alpha,\beta}E_{\alpha+\beta},
\end{equation}
$K_{\beta\beta}^{\alpha}=0$ also
$K_{\beta\gamma}^{\alpha}=N_{\beta,\gamma}$ if
$\beta+\gamma=\alpha$ and $K_{\beta\gamma}^{\alpha}=0$ if
$\beta+\gamma\neq\alpha$ in the root sense. By using \eqref{215}
the last equation in \eqref{213} can also be written as
\begin{equation}\label{220}
\begin{aligned}
d(e^{-\sigma}L\ast F)&=d\sigma\wedge e^{-\sigma}L\ast
F-\mathcal{G}_{0}^{T}\wedge
e^{-\sigma}L\ast F\\
\\
&\quad+2(L^{T})^{-1}\eta \,e^{-2\sigma}F\wedge e^{-\sigma}\ast G.
\end{aligned}
\end{equation}
The field equations of the dilatons and the axions for a generic
non-split scalar coset $G/K$ coupled to matter fields have been
derived in \cite{ker1,ker2,nej3}. Thus referring to
\cite{nej1,nej2,ker1,ker2,nej3} by varying the Lagrangian
\eqref{210} with respect to the axions $\{\chi^{m}\}$ and the
dilatons $\{\phi^{i}\}$ the corresponding field equations can be
given as
\begin{equation}\label{221}
\begin{aligned}
d(e^{\frac{1}{2}\gamma _{i}\phi ^{i}}\ast U^{\gamma
})&=-\frac{1}{2}\gamma _{j}e^{\frac{1}{2}\gamma _{i}\phi
^{i}}d\phi ^{j}\wedge \ast U^{\gamma }\\
\\
&\quad+\sum\limits_{\alpha -\beta =-\gamma }e^{\frac{1}{2} \alpha
_{i}\phi ^{i}}e^{\frac{1}{2}\beta _{i}\phi ^{i}}N_{\alpha ,-\beta
}U^{\alpha }\wedge \ast U^{\beta
},\\
\\
d(\ast d\phi ^{i})&=\frac{1}{2}\sum\limits_{\alpha\in\Delta_{nc}^{+}}^{}\alpha _{i}%
e^{\frac{1}{2}\alpha _{j}\phi ^{j}}U^{\alpha }\wedge e^{%
\frac{1}{2}\alpha _{j}\phi ^{j}}\ast U^{\alpha
}\\
\\
&\quad+2e^{-2\sigma}((H_{i})_{N}^{A}L_{M}^{N}L_{J}^{A})\ast
F^{M}\wedge F^{J}.
\end{aligned}
\end{equation}
Here the matrices $\{(H_{i})_{N}^{A}\}$ are the matrix
representatives of the generators $\{H_{i}\}$ in the
($N+1$)-dimensional representation we assume.
\section{Dualisation and the First-Order Formulation}

In this section we will apply the dualisation method of
\cite{julia2} to construct the coset formulation of the bosonic
sector of the $D=9$ matter coupled supergravity. We will also
obtain the first-order bosonic field equations as a consequence of
the coset formulation. Our formulation will be in parallel with
the ones in \cite{nej4,nej5}. We will derive the Lie superalgebra
which parameterizes the doubled or the dualized coset element so
that the doubled Cartan form induced by it realizes the
second-order field equations \eqref{213} and \eqref{221} by
satisfying the Cartan-Maurer equation. Likewise in
\cite{julia2,nej4,nej5} the doubled Cartan form of the dualized
coset element will obey a twisted self-duality equation which
results in the first-order bosonic field equations of the theory.
We start by assigning a generator for each bosonic field. The
original generators $\{K,V_{I},Y,H_{j},E_{m}\}$ will be coupled to
the fields $\{\sigma,A^{I},B,\phi^{j},\chi^{m}\}$ in the dualized
coset parametrization. We will also introduce a dual field for
each original field. The dual fields are
$\{\widetilde{\sigma},\widetilde{A}^{I},\widetilde{B},
\widetilde{\phi}^{j},\widetilde{\chi}^{m}\}$ where $\widetilde{B}$
is a five-form, $\{\widetilde{A}^{I}\}$ are six-forms and
$\{\widetilde{\sigma},\widetilde{\phi}^{j},\widetilde{\chi}^{m}\}$
are seven-forms. These dual fields are the Langrange multipliers
corresponding to the Bianchi identities of the relative original
fields \cite{pope}. For example $\widetilde{B}$ is the Langrange
multiplier which is coupled to the Bianchi identity of $B$ in the
construction of the Bianchi Lagrangian term which leads to the
first-order algebraic equations of motion etc. We will also assign
the dual generators
$\{\widetilde{K},\widetilde{V}_{I},\widetilde{Y},
\widetilde{H}_{j},\widetilde{E}_{m}\}$ to the dual fields
$\{\widetilde{\sigma},\widetilde{A}^{I},\widetilde{B},
\widetilde{\phi}^{j},\widetilde{\chi}^{m}\}$ so that they will
couple with each other in the parametrization of the dualized
coset element. The Lie superalgebra of the original and the dual
generators will have the $\mathbb{Z}_{2}$ grading so that the
generators will be odd if the corresponding potential is an odd
degree differential form and otherwise even \cite{julia2}. The
generators $\{V_{I},\widetilde{K},\widetilde{Y},
\widetilde{H}_{j},\widetilde{E}_{m}\}$ are odd generators and
$\{K,\widetilde{V}_{I},Y,H_{j},E_{m}\}$ are even ones.
Specifically the doubled coset element will be parameterized by a
differential graded algebra. This algebra is generated by the
differential forms and the generators we have introduced above.
The odd (even) generators behave like odd (even) degree
differential forms under this graded differential algebra
structure when they commute with the differential forms. The odd
generators obey the anti-commutation relations while the even ones
and the mixed couples obey the commutation relations.

To begin the construction of the coset formulation or the
non-linear realisation of the bosonic sector of the $D=9$ matter
coupled supergravity we propose the dualized coset element
\begin{equation}\label{31}
\nu^{\prime}=e^{\frac{1}{2}\phi^{j}H_{j}}e^{\chi^{m}E_{m}}e^{\sigma
K}e^{A^{I}V_{I}}e^{\frac{1}{2}BY}
e^{\frac{1}{2}\widetilde{B}\widetilde{Y}}e^{\widetilde{A}^{I}\widetilde{V}_{I}}e^{\widetilde{\sigma}
\widetilde{K}}e^{\widetilde{\chi}^{m}\widetilde{E}_{m}}e^{\frac{1}{2}\widetilde{\phi}^{j}\widetilde{H}_{j}}.
\end{equation}
Here $\nu^{\prime}$ is a map from the nine-dimensional spacetime
into a group which is presumably the rigid symmetry group of the
dualized Lagrangian. Likewise in \cite{julia2,nej4,nej5} we are
not interested in the group theoretical structure of the
non-linear realisation of the $D=9$ matter coupled supergravity
but rather in the Lie superalgebra which generates \eqref{31}.
Eventually this algebra also contains the information of the group
theoretical structure of the coset formulation
\cite{nej1,nej2,ker1,ker2,julia1,hel}. The local map
$\nu^{\prime}$ induces the doubled Cartan form
$\mathcal{G}^{\prime}$ on the nine-dimensional spacetime which can
be given as
\begin{equation}\label{32}
\mathcal{G}^{\prime}=d\nu^{\prime}\nu^{\prime-1}.
\end{equation}
From its construction the Cartan form \eqref{32} satisfies the
Cartan-Maurer equation
\begin{equation}\label{33}
d\mathcal{G}^{\prime}-\mathcal{G}^{\prime}\wedge\mathcal{G}^{\prime}=0.
\end{equation}
The standard dualisation procedure
\cite{julia2,nej1,nej2,nej3,nej4,nej5} requires the construction
of the Lie superalgebra of the original and the dual generators
such that they will lead us to the second-order equations of
motion \eqref{213} and \eqref{221} when the Cartan form \eqref{32}
is calculated and inserted in \eqref{33}. Therefore by using the
matrix formulas
\begin{equation}\label{34}
\begin{aligned}
de^{X}e^{-X}&=dX+\frac{1}{2!}[X,dX]+\frac{1}{3!}[X,[X,
dX]]+....,\\
\\
e^{X}Ye^{-X}&=Y+[X,Y]+\frac{1}{2!}[X,[X,Y]]+....,
\end{aligned}
\end{equation}
repeatedly we can calculate the Cartan form \eqref{32} in terms of
the desired unknown structure constants of the Lie superalgebra of
the original and the dual generators we have introduced. If we use
this calculated Cartan form in the Cartan-Maurer equation
\eqref{33} and then compare the result with the second-order
bosonic equations \eqref{213} and \eqref{221} of the $D=9$ matter
coupled supergravity we can read the unknown structure constants
and determine the Lie superalgebra structure which leads to the
coset formulation of the $D=9$ matter coupled supergravity. We
will not present the details of this calculation here, similar
calculations can be referred in \cite{nej3,nej4,nej5}. However we
should mention one point. Likewise in \cite{nej4,nej5} the scalar
sector which is coupled to the matter fields in \eqref{210} can be
treated separately on its own respect and one can make use of the
general results obtained in \cite{nej2,nej3} to derive the
commutation relations of the scalar generators and their duals.
Therefore consequently if we perform the above mentioned
calculation the comparison of \eqref{33} with the second-order
bosonic field equations \eqref{213} and \eqref{221} leads us to
the algebra structure whose commutation and the anti-commutation
relations can be given as
\begin{subequations}\label{35}
\begin{gather}
[K,V_{I}]=-V_{I}\quad,\quad[K,Y]=-2Y\quad,\quad[K,\widetilde{Y}]=2\widetilde{Y},\notag\\
\notag\\
[\widetilde{V}_{I},K]=-\widetilde{V}_{I}\quad,\quad\{V_{I},V_{J}\}=\eta_{IJ}Y\quad,\quad
[H_{l},V_{I}]=(H_{l})_{I}^{K}V_{K},\notag\\
\notag\\
[E_{m},V_{I}]=(E_{m})_{I}^{J}V_{J}\quad,\quad[V_{L},\widetilde{V}_{M}]=-\frac{2}{7}\delta_{LM}\widetilde{K}
-\underset{i}{\sum
}(H_{i})_{LM}\widetilde{H}_{i},\notag\\
\notag\\
\{V_{K},\widetilde{Y}\}=4\eta_{K}^{L}\,\widetilde{V}_{L}\quad,\quad[Y,\widetilde{Y}]=-\frac{16}{7}\widetilde{K}
\quad,\quad[H_{i},\widetilde{V}_{K}]=-(H_{i}^{T})_{K}^{M}\widetilde{V}_{M},\notag\\
\notag\\
[E_{\alpha},\widetilde{V}_{K}]=-(E_{\alpha}^{T})_{K}^{M}\widetilde{V}_{M},\notag\\
\notag\\
[H_{j},E_{\alpha }]=\alpha _{j}E_{\alpha }\quad ,\quad [E_{\alpha
},E_{\beta }]=N_{\alpha ,\beta }E_{\alpha+\beta},\notag\\
\notag\\
[H_{j},\widetilde{E}_{\alpha }]=-\alpha _{j}\widetilde{E}_{\alpha }\quad ,\quad [%
E_{\alpha },\widetilde{E}_{\alpha }]=\frac{1}{4}\overset{r}{\underset{j=1}{%
\sum }}\alpha _{j}\widetilde{H}_{j},\notag\\
\notag\\
[E_{\alpha },\widetilde{E}_{\beta }]=N_{\alpha ,-\beta }\widetilde{E}%
_{\gamma },\quad\quad\alpha -\beta =-\gamma,\;\alpha \neq
\beta.\tag{\ref{35}}
\end{gather}
\end{subequations}
Here $(E_{\alpha}^{T})_{K}^{M}$ and $(H_{i}^{T})_{K}^{M}$ are the
components of the transpose of the matrix representatives of the
generators $E_{\alpha}$ and $H_{i}$ respectively. The commutators
and the anti-commutators which are not listed in \eqref{35}
vanish. The commutation relations of the scalar generators and
their duals form up a subalgebra and they are already derived in
\cite{nej2} for a generic scalar coset. If $O$ represents the set
of the original generators and $\widetilde{D}$ the set of the dual
generators we conclude that the algebra constructed in \eqref{35}
obeys the general scheme
\begin{subequations}\label{36}
\begin{gather}
[O,\widetilde{D}\}\subset\widetilde{D}\quad,\quad [O,O\}\subset O,\notag\\
\tag{\ref{36}}\\
[\widetilde{D},\widetilde{D}\}=0.\notag
\end{gather}
\end{subequations}
Now that we have obtained the commutation and the anti-commutation
relations of the original and the dual generators we can calculate
the Cartan form $\mathcal{G}^{\prime}=d\nu^{\prime}\nu^{\prime-1}$
explicitly. From the definition of $\nu^{\prime}$ in \eqref{31} by
using the identities \eqref{34} effectively together with the
algebra structure given in \eqref{35} the calculation of the
Cartan form $\mathcal{G}^{\prime}$ yields
\begin{equation}\label{37}
\begin{aligned}
\mathcal{G}^{\prime}&=d\nu^{\prime}\nu^{\prime-1}\\
\\
&=\frac{1}{2}d\phi^{i}H_{i}+e^{\frac{1}{2}\alpha_{i}\phi^{i}}U^{\alpha}E_{\alpha}
+d\sigma K+e^{-\sigma}L_{I}^{K}dA^{I}V_{K}+\frac{1}{2}e^{-2\sigma}GY\\
\\
&\quad+\frac{1}{2}e^{2\sigma}d\widetilde{B}\widetilde{Y}+(-\frac{4}{7}B\wedge
d\widetilde{B}+ \frac{2}{7}A^{J}\wedge
d\widetilde{A}^{I}\delta_{IJ}+d\widetilde{\sigma})\widetilde{K}\\
\\
&\quad+\overset{r}{\underset{m=1}{\sum}}((e^{\mathbf{\Gamma}}e^{\mathbf{\Lambda}})_{j}^{m}
S^{j}+(H_{m})_{JI}A^{J}\wedge
d\widetilde{A}^{I}\\
\\
&\quad+\eta_{I}^{K} (H_{m})_{JK}A^{J}\wedge A^{I} \wedge
d\widetilde{B})\widetilde{H}_{m}+\underset{\alpha\in\Delta_{nc}^{+}}{\sum}(e^{\mathbf{\Gamma}}e^{\mathbf{\Lambda}})_{j}^{\alpha+r}
S^{j}\widetilde{E}_{\alpha}\\
\\
&\quad+(e^{\sigma}((L^{T})^{-1})_{K}^{L}d\widetilde{A}^{K}+2e^{\sigma}((L^{T})^{-1})_{K}^{L}
\,\eta_{I}^{K}\,A^{I}\wedge d\widetilde{B})\widetilde{V}_{L},
\end{aligned}
\end{equation}
where we have defined $r=$dim$\mathbf{h}_{k}$ and also
$S^{i}=\frac{1}{2}d\widetilde{\phi}^{i}$ for $ i=1,...,r$ whereas
$S^{\alpha+r}=d\widetilde{\chi}^{\alpha}$ for
$\alpha=1,...,$dim$\mathbf{n}_{k}$. We assume that we enumerate
the roots $\alpha\in\Delta_{nc}^{+}$ as
$\alpha=1,...,$dim$\mathbf{n}_{k}$. The matrices $\mathbf{\Gamma}$
and $\mathbf{\Lambda}$ have already been derived and given in
\cite{nej1,nej2,nej3,nej4,nej5} for general scalar cosets whose
definitions are also applicable to the scalar coset
$SO(N,1)/SO(N)$ of the matter coupled $D=9$ supergravity. We
should remark that in our entire formulation we use the Euclidean
signature metric to raise and lower the indices when needed for
notational purposes. Since the dualisation of the supergravity
theories is another manifestation of the Langrange multiplier
methods \cite{pope} the dualized formulation of the theory also
yields the first-order field equations in terms of the dual fields
\cite{julia2,nej3,nej4,nej5}. By obeying the twisted self-duality
equation
\begin{equation}\label{38}
\ast\mathcal{G}^{\prime}=\mathcal{SG}^{\prime},
\end{equation}
the Cartan form $\mathcal{G}^{\prime}$ calculated in \eqref{37}
leads us to the first-order field equations which can be obtained
by the local integration of the second-order equations of motion
\eqref{213} and \eqref{221} \cite{julia2}. Here $\mathcal{S}$ is a
pseudo-involution of the Lie superalgebra generated by the
original and the dual generators. In general $\mathcal{S}$ sends
the original algebra generators to their duals and the
dual ones to their original counterparts with a sign factor which is $(-1)^{p(%
D-p)+s}$ where $p$ is the degree of the field strength of the
corresponding field which the generator is coupled to and $s$ is
the signature of the spacetime metric \cite{julia2}. For our case
of the $D=9$ matter coupled supergravity the signature of the
spacetime is assumed to be $s=8$ \cite{d=9}. So the sign factor
becomes $(-1)^{p(9-p)+8}$. Thus the action of the
pseudo-involution $\mathcal{S}$ on the generators can be given as
\begin{subequations}\label{39}
\begin{gather}
\mathcal{S}Y=\widetilde{Y}\quad,\quad\mathcal{S}\widetilde{Y}=Y\quad,\quad
\mathcal{S}E_{m}=\widetilde{E}_{m}\quad,\quad\mathcal{S}\widetilde{E}_{m}=E_{m},\notag\\
\notag\\
\mathcal{S}K=\widetilde{K}\quad,\quad\mathcal{S}\widetilde{K}=K\quad,\quad
\mathcal{S}H_{i}=\widetilde{H}_{i}\quad,\quad\mathcal{S}\widetilde{H}_{i}=H_{i},\notag\\
\notag\\
\mathcal{S}V_{I}=\widetilde{V}_{I}\quad,\quad\mathcal{S}\widetilde{V}_{I}=V_{I}.\tag{\ref{39}}
\end{gather}
\end{subequations}
Equipped with the above definition of the pseudo-involution
$\mathcal{S}$ if we apply the twisted self-duality condition
\eqref{38} on the dualized Cartan form \eqref{37} we obtain the
first-order field equations as
\begin{subequations}\label{310}
\begin{gather}
\frac{1}{2}e^{-2\sigma}\ast G=\frac{1}{2}e^{2\sigma}d\widetilde{B},\notag\\
\notag\\
e^{-\sigma}\L_{J}^{I}\ast
dA^{J}=e^{\sigma}((L^{T})^{-1})_{J}^{I}d\widetilde{A}^{J}+2e^{\sigma}
((L^{T})^{-1})_{J}^{I}\,\eta_{K}^{J}\, A^{K}\wedge d\widetilde{B},\notag\\
\notag\\
\ast d\sigma=d\widetilde{\sigma}-\frac{4}{7}B\wedge d\widetilde{B}+\frac{2}{7}\delta_{IJ}A^{J}\wedge d\widetilde{A}^{I},\notag\\
\notag\\
e^{\frac{1}{2}\alpha_{i}\phi^{i}}(\mathbf{\Omega})_{l}^{\alpha}\ast
d\chi^{l}=(e^{\mathbf{\Gamma}}
e^{\mathbf{\Lambda}})_{j}^{\alpha+r}S^{j},\notag\\
\notag\\
\ast
d\phi^{m}=2(e^{\mathbf{\Gamma}}e^{\mathbf{\Lambda}})_{j}^{m}S^{j}+
2(H_{m})_{JI}A^{J}\wedge d\widetilde{A}^{I}+2\eta_{I}^{K}
(H_{m})_{JK}A^{J}\wedge A^{I} \wedge d
\widetilde{B}.\tag{\ref{310}}
\end{gather}
\end{subequations}
Taking the exterior derivative of the equations given in
\eqref{310} yields us the second-order field equations \eqref{213}
and \eqref{221} which are independent of the dual fields
naturally. This calculation is a straightforward one however two
remarks must be made: firstly one should make use of the
identities \eqref{27} and \eqref{28} which the coset elements obey
especially in the differentiation of the dilaton equation in
\eqref{310} to obtain the second-order dilaton equation in
\eqref{221}, secondly the general results of \cite{nej3} which
presents the dualisation and the first-order formulation of the
generic matter-coupled non-split scalar cosets may be adopted for
the differentiation of the dilaton and the axion equations in
\eqref{310} since these scalar fields parameterize the coset
$SO(N,1)/SO(N)$ where $SO(N,1)$ is a non-compact real form of a
semi-simple Lie group and thus $SO(N,1)/SO(N)$ is in general a
non-split scalar coset as we have mentioned before.

\section{Conclusion}
By introducing dual fields for the bosonic field content we have
performed the method of dualisation \cite{julia2} to obtain the
non-linear realisation of the bosonic sector of the $D=9$ matter
coupled supergravity \cite{d=9}. For the scalars of the coupling
vector multiplets we have given the symmetric space sigma model
formulation which is based on the $SO(N,1)/SO(N)$ coset manifold
in section two. In section three we have extended the non-linear
formulation of the scalars to the entire bosonic sector excluding
the graviton. We have doubled the bosonic field content by
defining dual fields. After introducing an algebra generator for
each field we have proposed a doubled coset element. Our objective
has been to construct the Lie superalgebra of the field generators
which parameterize the doubled coset element. We have derived the
necessary commutation and the anti-commutation relations of the
generators by calculating the dualized Cartan form due to the
dualized coset element in terms of the unknown structure constants
and by inserting it into the Cartan-Maurer equation then by
comparing the result with the second-order field equations which
have been given in section two. In summary we have derived the Lie
superalgebra which realizes the bosonic field equations of the
$D=9$ matter coupled supergravity by parameterizing a coset
element whose Cartan form yields the field equations in the
Cartan-Maurer equation. Thus this construction leads us to the
bosonic non-linear realisation of the $D=9$ matter coupled
supergravity. We have also obtained the first-order bosonic field
equations of the $D=9$ matter coupled supergravity as a twisted
self-duality equation \cite{julia2} of the dualized Cartan form.
Thus the dualisation method which is closely related to the
Lagrange multiplier construction \cite{pope} gives us also the
first-order equations of motion in terms of the dual fields.

In this work although we have derived the algebra structure which
parameterizes the dualized coset element we have neither shown an
attempt to identify the algebra nor worked on the group
theoretical construction of the coset. One can focus on the
structure of the algebra given in section three and inspect its
correspondence with the ones formulated in \cite{nej4} and
\cite{nej5}. We can also examine the local and the global
symmetries of the first-order field equations and the dualized
Cartan form and these symmetries can be linked to the coset
parametrization we have constructed. This analysis would give the
group theoretical construction of the dualized coset. The graviton
can also be included in the non-linear coset formulation presented
here. This would lead one to a non-linear realisation of the
theory in terms of the Kac-Moody algebras \cite{d=83,d=84,d=85}.
One can extend the non-linear coset formulation to include the
fermions as well. The knowledge of the global and the local
symmetries which can be extracted from the coset formulation
presented here will have a contribution to the understanding of
the symmetries of the ten dimensional heterotic string theory
since the $D=9$ matter coupled supergravity can be obtained as a
Kaluza-Klein descendant theory of the low energy limit of the ten
dimensional heterotic string \cite{tez21}. Such a symmetry
inspection would be improved by the more general non-linear
realisation of the matter coupled $D=9$ supergravity which would
include the gravity sector and whose formulation would be based on
a Kac-Moody symmetry group as in \cite{d=83,d=84,d=85}.

\end{document}